\documentclass[%
 reprint,aps,
 amsmath,amssymb,longbibliography,superscriptaddress]{revtex4-2}
\usepackage{booktabs}
\usepackage{tikz}
\usepackage{calc}
\usepackage{natbib}
\usepackage{booktabs}
\usepackage{longtable}
\usepackage{hyperref}
\hypersetup{
    colorlinks=true,
    linkcolor=black,
    filecolor=black,      
    urlcolor=black,
    citecolor=black,
    breaklinks
}

\usetikzlibrary{intersections}
\usetikzlibrary{decorations.pathmorphing}
\usetikzlibrary{calc}
\usetikzlibrary{shapes}

\usepackage{amsmath}

\begin{document}

\title{Correlators from Amplitubes}%

\author{Ross Glew}%
\email{r.glew@herts.ac.uk}
\affiliation{%
  Department of Physics, Astronomy and Mathematics, University of Hertfordshire, Hatfield, Hertfordshire, AL10 9AB, UK
}%

\begin{abstract}
Recently, the wavefunction coefficients for conformally coupled scalars in an FRW cosmology have been presented as a sum over amplitude-like functions known as {\it amplitubes}. In this work we extend this analysis to full {\it correlation functions}. Remarkably, the amplitube expansion of the correlator exhibits many vanishing contributions that are otherwise present in the expansion of the wavefunction. Moreover, while the wavefunction coefficients suffer from relative minus signs between terms, the surviving terms in correlation function do not. These observations point to a hidden simplicity in the structure of the correlation function compared to that of the wavefunction coefficient.
\end{abstract}

\maketitle
\section{Introduction}
In a cosmological setting, late-time observers have access to spatial {\it correlation functions} evaluated at a fixed time-slice. These correlators serve as the primary observables, from which one seeks to construct a cosmological history that explains the observed spatial patterns. Familiar from standard quantum mechanics, the correlators are obtained via application of the Born rule to the so called {\it wavefunction of the univserse} $\Psi[\Phi]$. 

In recent years, significant progress has been made towards understanding the wavefunction for conformally coupled scalars with polynomial interactions in an FRW cosmology. Much of the focus has been on computing the wavefunction coefficients on a graph-by-graph basis. This has lead to significant developments including: bootstrap approaches \cite{Arkani-Hamed:2018kmz,Jazayeri:2021fvk}, geometric formulations \cite{Arkani-Hamed:2017fdk,Benincasa:2019vqr,Benincasa:2024leu}, differential equations \cite{Arkani-Hamed:2023kig,Baumann:2024mvm,Baumann:2025qjx,De:2023xue,He:2024olr,De:2024zic,Grimm:2025zhv,Capuano:2025ehm,Qin:2024gtr,Hang:2024xas} and more \cite{AguiSalcedo:2023nds,Melville:2021lst,Goodhew:2020hob,Goodhew:2021oqg,De:2025bmf,Benincasa:2024ptf}. More recently, the sum over graphs has itself received a similar geometric treatment, mirroring developments for scattering amplitudes \cite{Arkani-Hamed:2017mur,Arkani-Hamed:2024jbp,Figueiredo:2025daa}. Of particular relevance to this work is a simple formula for the contribution of a fixed graph $G$ to the wavefunction coefficient, expressed as a sum over amplitude-like objects as \cite{Glew:2025ugf}
\begin{align}
\Psi_G = \sum_{I \subset E_G} (-1)^{|I|}  A_{G \setminus I},
\label{eq:wfc_int}
\end{align}
where the functions appearing on the right are referred to as {\it amplitubes} \cite{Glew:2025otn}. 

However, since the ultimate goal is to compute correlation functions, it stands to reason that we should deal with these objects directly. Therefore, the aim of this paper will be to find an amplitube expansion of the correlator itself, analogous to \eqref{eq:wfc_int} for the wavefunction coefficients. Our starting point will be a known formula for the contribution of a fixed graph $G$ to the correlator given by
\begin{align}
\langle G \rangle =\frac{ \mathcal{N}}{\prod_{e \in E_G} (2y_e)}  \sum_{I \subset E_G} \prod_{g \in \kappa_{G\setminus I}} (2\Psi_g).
\label{eq:corr_int}
\end{align}
At first site this formula suggests that the correlator is {\it more} complicated than the wavefunction coefficients, as it is expressed as a sum over entire wavefunction coefficients itself. However, when combined with \eqref{eq:wfc_int} we find this complexity is artificial. Somewhat miraculously, the factors of $2$ and relative minus signs conspire to cancel many terms in the sum, as well as removing all relative coefficients between surviving terms. After these simplifications, we find the resulting amplitube expansion of the correlator can be written as 
\begin{align}
\langle G \rangle = \frac{2\mathcal{N}}{\prod_{e \in E_G} (2y_e)}  \sum_{I \subset E_G}  \chi_{G/\bar{I}} A_{G \setminus I},
\end{align}
where $\chi_{G/\bar{I}}=1$ if the graph $G/\bar{I}$ is bipartite and vanishes otherwise. This formula reveals a hidden simplicity in the structure of the correlation function compared to that of the wavefunction coefficients. In fact, this simplicity was recently observed in \cite{Chowdhury:2023arc,Chowdhury:2025ohm,Donath:2024utn}, where it was shown that correlators can be obtained by ‘dressing-up’ flat-space amplitudes. Our formula, however, provides a combinatorial origin for these simplifications.

The remainder of the paper is organised as follows. In section~\ref{sec:rev} we review the definition of the wavefunction coefficients and correlation functions for our toy model of conformally coupled scalars. In section \ref{sec:wf} we detail the expansion of the wavefunction in terms of amplitubes. Finally, in section \ref{sec:corr} we present our main result, the amplitube expansion for the correlator. 
\section{Review}\label{sec:rev}
The focus of this work will be on a toy model of massless scalars in flat space with time dependent polynomial interactions, with action given by
\begin{align}
S = \int d^d x d \eta \left[ \frac{1}{2} \left( \partial \phi \right)^2 - \sum_{k \geq 3} \frac{\lambda_k(\eta)}{k!} \phi^k \right].
\end{align}
For a specific choice of $\lambda_k(\eta)$ this is equivalent to a theory of conformally coupled scalars with polynomial interactions in an FRW cosmology \cite{Arkani-Hamed:2017fdk}. However, for our purposes we focus on the flat-space case for which $\lambda_k(\eta)=const$. 

We will be interested in the set of {\it correlation functions} at fixed time $\eta^*=0$. As in standard quantum mechanics, these correlators are obtained via the Born rule as
\begin{align}
\langle \Phi(\vec{x}_1) \ldots \Phi(\vec{x}_n) \rangle = \frac{\int \mathcal{D}\Phi \Phi(\vec{x}_1) \ldots \Phi(\vec{x}_n) |\Psi[\Phi]|^2}{\int \mathcal{D}\Phi  |\Psi[\Phi]|^2},
\end{align}
where $\Phi(\vec{x})=\phi(\eta=0,\vec{x})$ denotes the boundary configuration of the field at $\eta=0$, and the mod square of the {\it flat-space wavefunction}, $|\Psi[\Phi]|^2$, serves as a probabliity distribution over the space of possible field configurations. 

The flat-space wavefunction is defined as the path integral over field configurations interpolating between the Bunch-Davies vacuum in the far past and the specified field configuration $\Phi(\vec{x})$ at fixed time in the future. Working in Fourier space, the wavefunction can be expanded as the following exponential
\begin{align}
\Psi = \exp \left\{ \sum_{n\geq 2} \int \prod_{i=1}^n d^d k_i \Psi_n[\vec{k}_i] \delta^d\left( \sum_{i=1}^n \vec{k}_i \right) \right\},
\end{align}
where the {\it wavefunction coefficients} $\Psi_n$ depend on the $n$ spatial momenta $\{\vec{k}_i\}$ associated to the field insertions on the late-time boundary. Following standard terminology we refer to the magnitude of the spatial momenta $|\vec{k}_i|$ as energies.

\subsection{Graph contributions}
\begin{figure}
\centering
\includegraphics[scale=0.75]{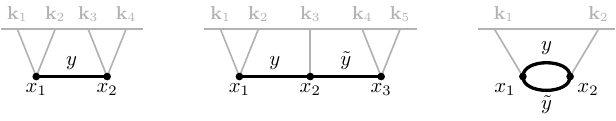}
\caption{Example graph contributions to the wavefunction coefficients.}
\label{figs:feyn}
\end{figure}

Both the wavefunction coeffcients and the correlator can be computed via a sum over Feynman graphs $\mathcal{G}$, examples of which are displayed in Fig.~\ref{figs:feyn}. Given a Feynman graph $\mathcal{G}$ it is useful to consider the truncated graph $G$ obtained from $\mathcal{G}$ by deleting all external (grey) edges which extend to the boundary. We denote the contribution to (the integrand of) the wavefunction coefficients and the correlator from such a truncated graph by $\Psi_G$ and $\langle G \rangle$ respectively. 

Since we are dealing with conformally coupled scalars it is natural to introduce a set of variables associated to the vertices and edges of the graph. To the vertices we assign the variables $x_v$ given by the sum over all external energies flowing into the vertex. To the edges we assign the variable $y_e$ given by the energy associated to the corresponding internal edge of $\mathcal{G}$. 

As an example consider the two tree-level processes in Fig.~\ref{figs:feyn}. The first is a four-point $s$-channel contribution which has graph variables given by
\begin{align}
x_1 &= |\vec{k}_1|+|\vec{k}_2|,  \quad \quad \quad x_2 = |\vec{k}_3|+ |\vec{k}_4|, \notag \\ 
\quad y &= |\vec{k}_1+\vec{k}_2|+|\vec{k}_3+\vec{k}_4|.
\end{align}
Similarly, for the five-point process we have the following graph variables
\begin{align}
x_1 &= |\vec{k}_1|+|\vec{k}_2|,  \quad x_2 = |\vec{k}_3|,  \quad x_2 = |\vec{k}_3|+|\vec{k}_4|, \notag \\
y &= |\vec{k}_1+\vec{k}_2|=|\vec{k}_3+\vec{k}_4+\vec{k}_5|, \notag \\
\tilde{y} &= |\vec{k}_4+\vec{k}_5|=|\vec{k}_1+\vec{k}_2+\vec{k}_3|.
\end{align}
Furthermore, the wavefunction coefficients/correlator are naturally expressed in terms of variables associated to connected subgraphs or rather {\it tubes} of $G$. To each {\it tube} $t$ we associate a variable $H_t$ given by the sum over $x_v$ for all $v \in t$ plus the sum over $y_e$ for each edge $e$ crossed by the tube counted with multiplicity, that is 
\begin{align}
H_t = \sum_{v \in t} x_v + \sum_{e \text{ cross } t} y_e.
\end{align}
For the graphs depicted in Fig.~\ref{figs:feyn} we have the following tube variables: for the two-chain
\begin{align}
\raisebox{-0.05cm}{\includegraphics[scale=0.8]{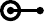} }= x_1+y, \quad \raisebox{-0.05cm}{\includegraphics[scale=0.8]{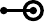}}=x_2+y, \quad \raisebox{-0.05cm}{\includegraphics[scale=0.8]{figs/pt2_t5.pdf}}=x_1+x_2,
\end{align}
for the three-chain,
\begin{align}
\raisebox{-0.02cm}{\includegraphics[scale=0.6]{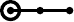}} &=x_1+y, \quad \raisebox{-0.02cm}{\includegraphics[scale=0.6]{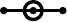}} =x_2+y+\tilde{y}, 
\quad \raisebox{-0.02cm}{\includegraphics[scale=0.6]{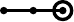}} =x_3+\tilde{y},  \notag \\
\raisebox{-0.11cm}{\includegraphics[scale=0.6]{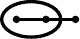}} &=x_1+x_2+\tilde{y},  \quad \raisebox{-0.11cm}{\includegraphics[scale=0.6]{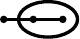}} =x_2+x_3+y, \notag \\
\raisebox{-0.11cm}{\includegraphics[scale=0.6]{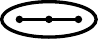}} &=x_1+x_2 + x_3. 
\end{align}
for the two-cycle,
\begin{align}
\raisebox{-0.16cm}{\includegraphics[scale=0.6]{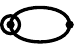}} &=x_1+y+\tilde{y},   &&\ \  \raisebox{-0.16cm}{\includegraphics[scale=0.6]{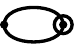}}=x_2+y+\tilde{y}, \notag\\
\raisebox{-0.16cm}{\includegraphics[scale=0.6]{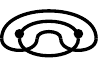}} &=x_1+x_2+2\tilde{y},  && \raisebox{-0.16cm}{\includegraphics[scale=0.6]{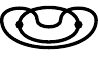}}=x_1+x_2+2y, \notag\\
\raisebox{-0.16cm}{\includegraphics[scale=0.6]{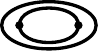}} &=x_1+x_2.
\end{align}
Where we have simply depicted the tube in place of using the notation $H_t$. 
\section{Wavefunction coefficients}
\label{sec:wf}
In this section we present an expansion of the wavefunction coefficients expressed as a sum over amplitude-like objects referred to as  {\it amplitubes}.
\subsection{Amplitubes}
Let $G=(E_G,V_G)$ be a graph. To define the {\it amplitubes}, we focus on the subset of tubes formed as induced subgraphs on a subset of vertices of $G$. It is convenient to label these tubes by the subset of vertices $t \subset V_G$ they contain. The set of all such vertex induced tubes is denoted by $T_G$.  We say two tubes $t,t' \in T_G$:
\begin{enumerate}
\item {\it intersect} if $t \cap t' \neq \emptyset$ and  $t \not\subset t'$ and $t' \not\subset t$,
\item{\it touch} if $t \cap t' = \emptyset$ and $t \cup t' \in T_G$,
\item are {\it compatible} if they do not intersect or touch. 
\end{enumerate}
A tubing $\tau$ is then defined to be a {\it maximal} set of pairwise compatible tubes of $T_G$. The set of tubings will be denoted by $\Gamma_G$.  

Finally, with this terminology introduced, the {\it amplitube} associated to the graph $G$ is defined as
\begin{align}
A_G = \sum_{\tau \in \Gamma_G} \prod_{t \in \tau} \frac{1}{H_t}.
\end{align}
For a more detailed discussion of these functions as well as their relation to geometry see \cite{devadoss2009realization,Glew:2025otn}.
\subsection{Wavefunction coefficients from amplitubes}
Having defined the amplitubes we can immediately write down the desired formula for the graph contribution to the wavefunction coefficient as \cite{Glew:2025ugf}
\begin{align}
\Psi_G =\sum_{I \subset E_G} (-1)^{|I|}  A_{G \setminus I}.
\label{eq:wfc_main}
\end{align}
In the above we have introduced the notation
\begin{align}
A_{G \setminus I} = \prod_{g \in \kappa_{G \setminus I}} A_g,
\end{align}
where $G\setminus I$ denotes the graph $G$ with edge set $I$ removed and $\kappa_{G\setminus I}$ its connected components. This formula is best understood through examples. 

In what follows we depict the deleted edge set $I$ by dotted edges. Take the two-chain, the wavefunction coefficient can be written as 
\begin{align} 
\Psi_{\includegraphics[scale=0.7]{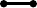}}&=A_{\includegraphics[scale=0.7]{figs/pt2.pdf}}-A_{\includegraphics[scale=0.7]{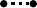}},
\end{align}
where the amplitubes are given by
\begin{align} 
A_{\includegraphics[scale=0.7]{figs/pt2.pdf}}&=\frac{1}{\includegraphics[scale=0.7]{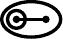}}+\frac{1}{\includegraphics[scale=0.7]{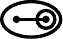}}, \quad A_{\includegraphics[scale=0.7]{figs/pt2_br.pdf}}&=\frac{1}{\includegraphics[scale=0.7]{figs/pt2_t4.pdf}} \times \frac{1}{\includegraphics[scale=0.7]{figs/pt2_t5.pdf}}.
\end{align}
Next, consider the three-chain, whose wavefunction coefficient is given by
\begin{align}
\Psi_{\raisebox{0.04cm}{\includegraphics[scale=0.7]{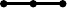}}}  = A_{\raisebox{0.04cm}{\includegraphics[scale=0.7]{figs/pt3}}}-A_{\raisebox{0.04cm}{\includegraphics[scale=0.7]{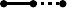}}}-A_{\raisebox{0.04cm}{\includegraphics[scale=0.7]{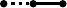}}}+A_{\raisebox{0.04cm}{\includegraphics[scale=0.7]{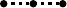}}}.
\end{align}
The amplitubes appearing in the above formula are expressed as
\begin{align}
A_{\raisebox{0.04cm}{\includegraphics[scale=0.55]{figs/pt3.pdf}}} &=\frac{1}{\raisebox{0.04cm}{\includegraphics[scale=0.55]{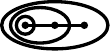}}}+\frac{1}{\raisebox{0.04cm}{\includegraphics[scale=0.55]{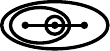}}}+\frac{1}{\raisebox{0.04cm}{\includegraphics[scale=0.55]{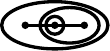}}}+\frac{1}{\raisebox{0.04cm}{\includegraphics[scale=0.55]{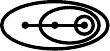}}}+\frac{1}{\raisebox{0.04cm}{\includegraphics[scale=0.55]{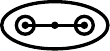}}}, \notag \\
A_{\raisebox{0.04cm}{\includegraphics[scale=0.55]{figs/pt3_br2.pdf}}} &=\left(\frac{1}{\raisebox{0.04cm}{\includegraphics[scale=0.55]{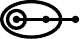}}} +\frac{1}{\raisebox{0.04cm}{\includegraphics[scale=0.55]{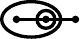}}}\right) \times \frac{1}{\raisebox{0.04cm}{\includegraphics[scale=0.55]{figs/pt3_t5_r.pdf}}} , \notag \\
A_{\raisebox{0.04cm}{\includegraphics[scale=0.55]{figs/pt3_br1.pdf}}} &=\frac{1}{\raisebox{0.04cm}{\includegraphics[scale=0.55]{figs/pt3_t5_l.pdf}}} \times \left(\frac{1}{\raisebox{0.04cm}{\includegraphics[scale=0.55]{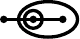}}} +\frac{1}{\raisebox{0.04cm}{\includegraphics[scale=0.55]{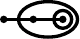}}} \right) , \notag \\
A_{\raisebox{0.04cm}{\includegraphics[scale=0.55]{figs/pt3_br12.pdf}}} &=\frac{1}{\raisebox{0.04cm}{\includegraphics[scale=0.55]{figs/pt3_t5_l.pdf}}} \times \frac{1}{\raisebox{0.04cm}{\includegraphics[scale=0.55]{figs/pt3_t5_m.pdf}}} \times \frac{1}{\raisebox{0.04cm}{\includegraphics[scale=0.65]{figs/pt3_t5_r.pdf}}}.
\end{align}
Finally, consider the two-cycle, with wavefunction coefficient given by
\begin{align}
\Psi_{\raisebox{-0.2cm}{\includegraphics[scale=0.5]{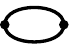}}}  =  A_{\includegraphics[scale=0.5]{figs/cycle_dec_4}}-A_{\includegraphics[scale=0.5]{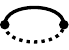}}-A_{\includegraphics[scale=0.5]{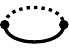}}+A_{\includegraphics[scale=0.5]{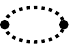}}.
\label{eq:loop_wf}
\end{align}
With the amplitubes defined as 
\begin{align}
A_{\includegraphics[scale=0.5]{figs/cycle_dec_4}}&=\frac{1}{\includegraphics[scale=0.6]{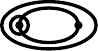}}+\frac{1}{\includegraphics[scale=0.6]{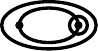}}, && A_{\includegraphics[scale=0.5]{figs/cycle_dec_2}}= \frac{1}{\includegraphics[scale=0.6]{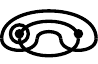}}+\frac{1}{\includegraphics[scale=0.6]{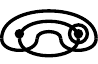}}, \notag \\
A_{\includegraphics[scale=0.5]{figs/cycle_dec_3}}&=\frac{1}{\includegraphics[scale=0.6]{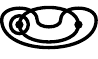}}+ \frac{1}{\includegraphics[scale=0.6]{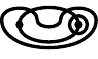}}, &&A_{\includegraphics[scale=0.5]{figs/cycle_dec_1}}=\frac{1}{\includegraphics[scale=0.6]{figs/cycle_tube_5l}} \times \frac{1}{\includegraphics[scale=0.6]{figs/cycle_tube_5r}} . \notag 
\end{align}
It is important to note that we have treated the amplitubes associated to subgraphs as being embedded in the parent graph $G$. This is to ensure that we obtain the correct tube variables $H_t$.
\section{Correlators from amplitubes}
\label{sec:corr}
Our aim now is to derive a similar formula for the graph contribution to the correlator, starting from the following expression \cite{Figueiredo:2025daa}
\begin{align}
\langle G \rangle =\frac{ \mathcal{N}}{\prod_{e \in E_G} (2y_e)}  \sum_{I \subset E_G} 2^{\kappa_{G \setminus I}} \Psi_{G \setminus I},
\label{eq:corr_main}
\end{align}
where the normalisation factor $\mathcal{N}$ is given by the product over $2|\vec{k}_i|$ for each external energy of the graph and again we have introduced the notation 
\begin{align}
\Psi_{G \setminus I}=\prod_{g \in \kappa_{G\setminus I}} \Psi_g.
\end{align}
To find the amplitube expansion of the correlator we simply combine \eqref{eq:wfc_main} with \eqref{eq:corr_main}. Before doing so in general, let us first consider how the examples of the previous section uplift to full correlation functions. The correlator for the two-chain is expressed in terms of wavefunction coefficients as 
\begin{align}
\langle \raisebox{0.04cm}{\includegraphics[scale=0.7]{figs/pt2.pdf}} \rangle = \frac{\mathcal{N}}{(2y)}\left(  2\Psi_{\includegraphics[scale=0.7]{figs/pt2.pdf}}+4\Psi_{\includegraphics[scale=0.7]{figs/pt2_br.pdf}} \right).
\label{eq:two_pt_corr}
\end{align}
The terms appearing above can be expanded in terms of amplitubes as  
\begin{align} 
\Psi_{\includegraphics[scale=0.7]{figs/pt2.pdf}}&= A_{\includegraphics[scale=0.7]{figs/pt2.pdf}}-A_{\includegraphics[scale=0.7]{figs/pt2_br.pdf}},  &&\Psi_{\includegraphics[scale=0.7]{figs/pt2_br.pdf}}=A_{\includegraphics[scale=0.7]{figs/pt2_br.pdf}} .
\end{align}
Substituting this into \eqref{eq:two_pt_corr} we find the following expression for the correlator
\begin{align}
\langle \raisebox{0.04cm}{\includegraphics[scale=0.7]{figs/pt2.pdf}} \rangle =\frac{2\mathcal{N}}{(2y)} \left( A_{\includegraphics[scale=0.7]{figs/pt2.pdf}}+A_{\includegraphics[scale=0.7]{figs/pt2_br.pdf}} \right).
\end{align}
Continuing our tree-level examples, the correlator for the three-chain is given by
\begin{align}
\langle \raisebox{0.04cm}{\includegraphics[scale=0.65]{figs/pt3.pdf}} \rangle  \propto 2\Psi_{\raisebox{0.04cm}{\includegraphics[scale=0.65]{figs/pt3}}}+4\Psi_{\raisebox{0.04cm}{\includegraphics[scale=0.65]{figs/pt3_br2.pdf}}}+4\Psi_{\raisebox{0.04cm}{\includegraphics[scale=0.65]{figs/pt3_br1.pdf}}}+8\Psi_{\raisebox{0.04cm}{\includegraphics[scale=0.65]{figs/pt3_br12.pdf}}},
\label{eq:three_corr}
\end{align}
where the wavefunction coefficients appearing on the right can be expanded as
\begin{align}
\Psi_{\raisebox{0.04cm}{\includegraphics[scale=0.7]{figs/pt3.pdf}}}  &= A_{\raisebox{0.04cm}{\includegraphics[scale=0.7]{figs/pt3}}}-A_{\raisebox{0.04cm}{\includegraphics[scale=0.7]{figs/pt3_br2.pdf}}}-A_{\raisebox{0.04cm}{\includegraphics[scale=0.7]{figs/pt3_br1.pdf}}}+A_{\raisebox{0.04cm}{\includegraphics[scale=0.7]{figs/pt3_br12.pdf}}}, \notag \\
\Psi_{\raisebox{0.04cm}{\includegraphics[scale=0.7]{figs/pt3_br2.pdf}}} &=A_{\raisebox{0.04cm}{\includegraphics[scale=0.7]{figs/pt3_br2.pdf}}} -A_{\raisebox{0.04cm}{\includegraphics[scale=0.7]{figs/pt3_br12.pdf}}}, \notag \\
\Psi_{\raisebox{0.04cm}{\includegraphics[scale=0.7]{figs/pt3_br1.pdf}}} &=A_{\raisebox{0.04cm}{\includegraphics[scale=0.7]{figs/pt3_br1.pdf}}} -A_{\raisebox{0.04cm}{\includegraphics[scale=0.7]{figs/pt3_br12.pdf}}},  \notag \\
\Psi_{\raisebox{0.04cm}{\includegraphics[scale=0.7]{figs/pt3_br12.pdf}}} &=A_{\raisebox{0.04cm}{\includegraphics[scale=0.7]{figs/pt3_br12.pdf}}}.
\end{align}
Combining this with \eqref{eq:three_corr} we find the following formula for the correlator
\begin{align}
\langle \raisebox{0.04cm}{\includegraphics[scale=0.65]{figs/pt3.pdf}} \rangle  \propto  \left( A_{\raisebox{0.04cm}{\includegraphics[scale=0.65]{figs/pt3}}}+A_{\raisebox{0.04cm}{\includegraphics[scale=0.65]{figs/pt3_br2.pdf}}}+A_{\raisebox{0.04cm}{\includegraphics[scale=0.65]{figs/pt3_br1.pdf}}}+A_{\raisebox{0.04cm}{\includegraphics[scale=0.65]{figs/pt3_br12.pdf}}}\right), 
\end{align}
with the proportionality factor given by $\frac{2\mathcal{N}}{(2y)(2\tilde{y})}$. Remarkably, in both cases the correlator can be obtained from the wavefunction coefficients, up to an overall factor, by removing all minus signs in \eqref{eq:wfc_main}, this statement is valid for all tree-level graphs. The amplitube expansion therefore uncovers a simple relation between the correlator and the wavefunction coefficients otherwise obscured by previous representations. 

At loop-level additional simplifications occur, to see why consider the two-cycle. The correlator in this case is given by
\begin{align}
\langle \raisebox{-0.17cm}{\includegraphics[scale=0.6]{figs/cycle_dec_4}} \rangle  \propto 2\Psi_{\includegraphics[scale=0.5]{figs/cycle_dec_4}}+2\Psi_{\includegraphics[scale=0.5]{figs/cycle_dec_2}}+ 2\Psi_{\includegraphics[scale=0.5]{figs/cycle_dec_3}}+ 4\Psi_{\includegraphics[scale=0.5]{figs/cycle_dec_1}}.
\end{align}
The wavefunction coefficents appearing in the expansion of the correlator are given by
\begin{align}
\Psi_{\raisebox{-0.2cm}{\includegraphics[scale=0.5]{figs/cycle_dec_4}}}  &=  A_{\includegraphics[scale=0.5]{figs/cycle_dec_4}}-A_{\includegraphics[scale=0.5]{figs/cycle_dec_2}}-A_{\includegraphics[scale=0.5]{figs/cycle_dec_3}}+A_{\includegraphics[scale=0.5]{figs/cycle_dec_1}}, \notag\\
\Psi_{\includegraphics[scale=0.5]{figs/cycle_dec_2}}&=  A_{\includegraphics[scale=0.5]{figs/cycle_dec_2}}-A_{\includegraphics[scale=0.5]{figs/cycle_dec_1}} , \notag \\
\Psi_{\includegraphics[scale=0.5]{figs/cycle_dec_3}}&=  A_{\includegraphics[scale=0.5]{figs/cycle_dec_3}}-A_{\includegraphics[scale=0.5]{figs/cycle_dec_1}} , \notag \\
\Psi_{\includegraphics[scale=0.5]{figs/cycle_dec_1}}&=A_{\includegraphics[scale=0.5]{figs/cycle_dec_1}}.
\end{align}
Substituting this into the expression for the correlator we find
\begin{align}
\langle \raisebox{-0.17cm}{\includegraphics[scale=0.6]{figs/cycle_dec_4}} \rangle =  \frac{2 \mathcal{N}}{(2 y)(2\tilde{y})} \left( A_{\includegraphics[scale=0.5]{figs/cycle_dec_4}}+A_{\includegraphics[scale=0.5]{figs/cycle_dec_1}} \right).
\end{align}
Comparing to \eqref{eq:loop_wf} we observe a new phenomenon, terms present in the wavefunction now vanish in the correlator. 

To further exhibit the remarkable simplifications which occur at loop level we can compare the wavefunction coefficient for the graph
\begin{align}
 \Psi_{\raisebox{-0.17cm}{\includegraphics[scale=0.5]{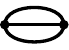}}} &=  A_{\includegraphics[scale=0.5]{figs/sunrise_1}}-A_{\includegraphics[scale=0.5]{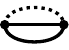}}-A_{\includegraphics[scale=0.5]{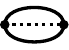}}-A_{\includegraphics[scale=0.5]{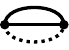}} \notag \\
&+A_{\includegraphics[scale=0.5]{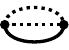}}+A_{\includegraphics[scale=0.5]{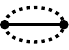}}+A_{\includegraphics[scale=0.5]{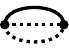}}-A_{\includegraphics[scale=0.5]{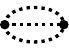}},
\end{align}
with the corresponding expression for the correlator
\begin{align}
\langle \raisebox{-0.17cm}{\includegraphics[scale=0.6]{figs/sunrise_1}} \rangle =  \frac{2 \mathcal{N}}{(2 y_1)(2y_2)(2y_3)} \left( A_{\includegraphics[scale=0.5]{figs/sunrise_1}}+A_{\includegraphics[scale=0.5]{figs/sunrise_8}} \right).
\end{align}
As these examples demonstrate, despite the complications of \eqref{eq:corr_main}, we find the correlator is in fact {\it simpler} than the wavefunction coefficient when expressed in terms of amplitubes! This is consistent with earlier observations presented in \cite{Chowdhury:2023arc,Chowdhury:2025ohm,Donath:2024utn}, where the simplicity of cosmological correlators and their relation to flat-space amplitudes was discussed.

In general, it is not hard to convince oneself that the coefficient in front of a given amplitube is given by
\begin{align}
\langle G \rangle = \frac{\mathcal{N}}{\prod_{e \in E_G} (2y_e)}  \sum_{I \subset E_G}  \sum_{J   \subset I} (-1)^{I\setminus J} 2^{\kappa_{G  \setminus J}}A_{G \setminus I}.
\end{align}
Remarkably, the coefficient multiplying the amplitube is a well known graph invariant which counts the number of two-colorings of the graph $G$ with edge set $\bar{I} = E_G \setminus I$ contracted, denoted by $G/\bar{I}$. Therefore, we arrive at the final form of the correlator given by  
\begin{align}
\langle G \rangle = \frac{2\mathcal{N}}{\prod_{e \in E_G} (2y_e)}  \sum_{I \subset E_G}  \chi_{G/\bar{I}} A_{G \setminus I},
\label{eq:main}
\end{align}
where we have defined
\begin{align}
\chi_{G/\bar{I}} \equiv \begin{cases} 1  & \text{if } G/\bar{I} \text{ bipartite}, \\ 0 & \text{ otherwise}.\end{cases}
\end{align}
We remind the reader that a graph is bipartite if its vertices can be partitioned into two disjoint sets such that no edge connects vertices within the same set.
 
\section{Conclusion}
In this paper we have provided a novel expression for correlators of conformally coupled scalars as an expansion over amplitubes, mimicking a similar expansion of the wavefunction coefficients presented in \cite{Glew:2025otn}. We have argued that this formula reveals a hidden simplicty of the correlator, as compared to the wavefunction coefficients, due to the cancellation of terms and tidying up of relative minus signs. 

The amplitube expansion of the wavefunction has a geometric interpretation as the triangulation of cosmological polytopes \cite{Arkani-Hamed:2017fdk}. It would be interesting to see if a similar geometric interpretation can be given to \eqref{eq:main} and to make connection with recently discovered geometries for the full correlator \cite{Arkani-Hamed:2024jbp,Figueiredo:2025daa}.

As a final speculative remark, we find it intriguing that a bipartite constraint naturally appears in the expression for the correlator, while flat-space amplitudes themselves can be formulated using bipartite graphs \cite{Arkani-Hamed:2012zlh}. It would be worthwhile to investigate whether these two occurrences of bipartiteness are fundamentally connected or merely coincidental. 

\acknowledgments
We would like to thank Arthur Lipstein and Chandramouli Chowdhury for bringing to our attention references on the subtle simplicity of cosmological correlators.
\bibliography{cor_v2}{} \bibliographystyle{unsrtnat}
\end{document}